\input{aipcheck}

\documentclass[
  ,draft            
  ]
  {aipproc}

\layoutstyle{6x9}

\def\gapp{\ifmmode\stackrel{>}{_{\sim}}\else$\stackrel{>}{_{\sim}}$\fi}

\begin{document}

\title{Searching for a Pulsar in SN1987A}

\classification{97.60.G6,97.60.Jd,98.38.Mz,98.56.Si}
\keywords      {supernovae:individual(SN 1987A) --- stars:neutron ---
pulsars:general}

\author{R N Manchester}{
  address={Australia Telescope National Facility, CSIRO, PO Box 76, Epping NSW 1710, Australia}
}

\begin{abstract}
SN 1987A offered a unique opportunity to detect a pulsar at the very
beginning of its life and to study its early evolution. Despite many
searches at radio and optical wavelengths, no pulsar has yet been
detected. Details of a recent search using the Parkes radio telescope
are given. Limits on the X-ray, optical and radio luminosity of a point
source at the centre of SN 1987A place limits on the properties of a
central neutron star. However, neither these nor the pulsar limits
preclude the presence of a relatively slowly rotating neutron star
($P\gapp 100$~ms) with a moderate surface dipole magnetic field in SN
1987A. Galactic studies suggest that a significant fraction of pulsars
are born with parameters in this range. In view of this, continued
searches for a pulsar in SN 1987A are certainly justified.
\end{abstract}

\maketitle


\section{Introduction}

As by far the nearest supernova to be observed in the modern era, SN
1987A offered a unique opportunity to search for a pulsar at the very
beginning of its life. The detection of a neutrino burst at the time
of the SN \cite{hkk+87,bbb+87} was good evidence that a neutron star
was formed at the time of the explosion. If such a pulsar were found,
it would of course be of immense interest to study its properties at
such an early stage of its life. 

The Crab pulsar remains the youngest known radio pulsar, with a
characteristic age, $\tau_c = P/(2\dot P)$, where $\dot P$ is the first
time derivative of the period, of 1240 years and a true age of about
950 years.\footnote{Pulsar parameters quoted in this paper have been
obtained from the ATNF Pulsar Catalogue
\url{http://www.atnf.csiro.au/research/pulsar/psrcat}. 
Original references are given in the catalogue.} Three apparently
younger pulsars are known, all detected only at X-ray or $\gamma$-ray
wavelengths: PSRs J1808$-$2024, J1846$-$0258 and J1907+0919. The first
and third are ``soft gamma-ray repeaters'', with long periods but very
large spin-down rates, giving them characteristic ages of just 280
years and 1050 years respectively. PSR J1846$-$0258 is an X-ray pulsar
lying at the center of the supernova remnant (SNR) Kes 75 with a
characteristic age of about 720 years.

\begin{figure}
\includegraphics[width=110mm]{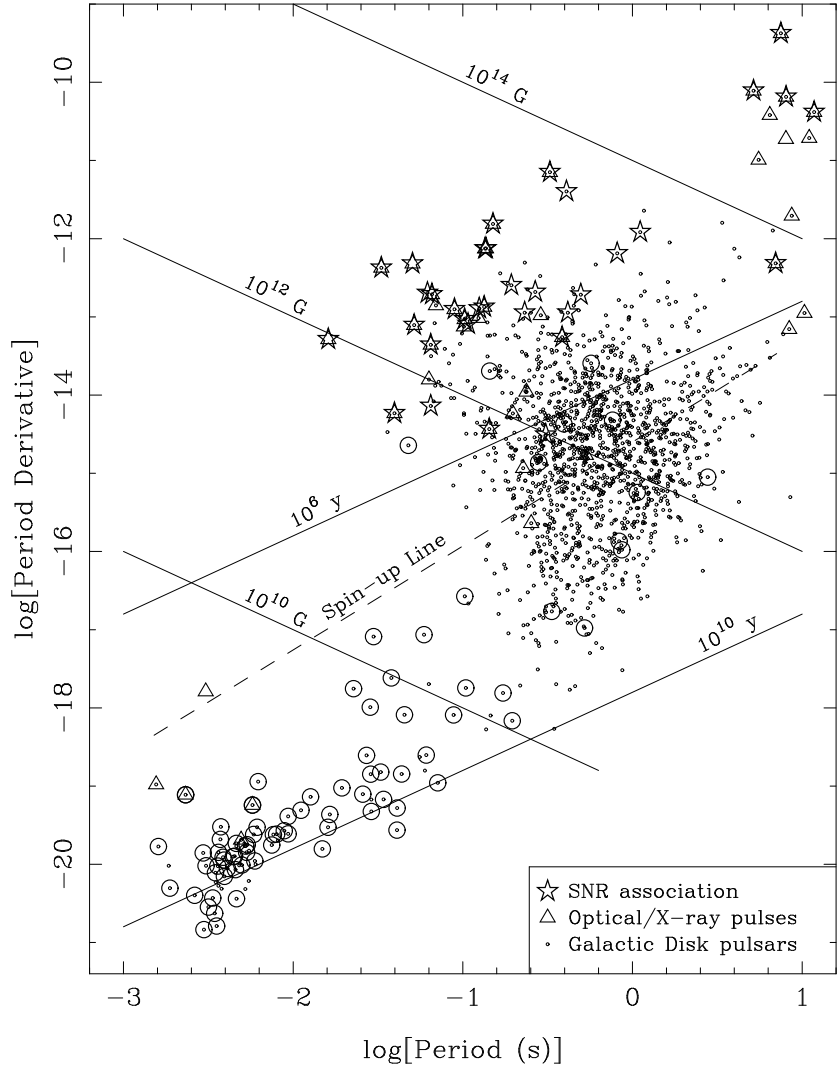}
\caption{Observed distribution of pulsars in the period -- period
  derivative plane. Lines of constant characteristic age and
  surface-dipole magnetic field strength are marked as is the limiting
  period for spin-up by accretion from a binary companion. Pulsars for
  which there is a plausible association with a supernova remnant and
  which emit at optical and higher frequencies are
  indicated.}\label{fg:ppdot} 
\end{figure}

Fig.~\ref{fg:ppdot} shows the observed distribution of pulsars in the
$P - \dot P$ plane. Anomalous X-ray pulsars (AXPs) and SGRs are
located in the top-right corner of the diagram with strong implied
surface dipole magnetic fields, $B_S \propto (P\dot P)^{1/2}$, and
long periods. There are about 35 plausible, or at least possible,
associations of pulsars with supernova remnants and these are all
relatively young pulsars, mostly with $\tau_c < 10^5$
years. Non-thermal pulsations in the optical, X-ray or $\gamma$-ray
bands are observed from about 45 pulsars and, apart from a few
millisecond pulsars, these pulsars are relatively young. Of the 35
pulsars in the catalogue with characteristic age less than 20,000
years, 21 are detected in one or more of the high-energy bands. Only
four or five pulsars have detectable non-thermal pulsations at optical
wavelengths \citep{kmmh03}, but these include two of the youngest
known pulsars, the Crab pulsar and PSR B0540$-$69. Furthermore, PSR
B0540$-$69 is located in the Large Magellanic Cloud, not far from SN
1987A. One other young pulsar, PSR J0537$-$6910, which has a period of
just 16 ms, is also located in the Large Magellanic Cloud close to SN
1987A. This pulsar is detected only at X-ray wavelengths
and is associated with the SNR N157B.

The way pulsars move on the $P - \dot P$ diagram is determined by the
time-dependence of the rotational braking torque. For many possible
braking mechanisms, the torque can be expressed as a power law
\begin{equation}\label{eq:torque}
N = -K \nu^n
\end{equation}
where $K$ is a constant, $\nu = 1/P$ and $n$ is known as the braking
index. For braking by reaction to emission of magnetic-dipole
radiation, $K \propto B_s^2$ where $B_s$ is the dipole magnetic field
strength at the surface of the neutron star, and $n=3$. The observed
value of $n$ is dominated by the secular braking in only the youngest
known pulsars and values of $n$ between 1.4 and 3.0 have been
measured. If the true age of pulsar, $\tau$, is known or can be estimated, for
example, from an association with a SNR, and the braking index
$n$ is assumed to be constant, the pulse period at birth can be
estimated from 
\begin{equation}
P_0 = P\left[1 - \frac{(n-1)}{2}\frac{\tau}{\tau_c}\right]^{1/(n-1)}
\end{equation}
where P is the current period. There are 12 SNR -- pulsar associations
where a reasonably reliable SNR age can be
estimated \citep{wlz06,tl06,nr07} leading to estimates of birth
period which range from $<11$~ms for PSR J0538$-$6910 to about 420~ms
for PSR J1210-5226, associated with the large shell SNR
G296.5+10.0. For the Crab pulsar, the estimated birth period is about
19 ms. 

A statistical study of the pulsars detected in the Parkes
Multibeam Pulsar Survey (PMPS) \citep{mlc+01} taking into account the survey
selection effects \citep{vml+04} suggests that 40\% of pulsars are born with periods
in the range 100 -- 500 ms. From an analysis of pulsars
detected in the PMPS and the Swinburne intermediate-latitude pulsar
survey, \citet{fk06} obtained consistent results with a gaussian
initial-period distribution centred at 300~ms and with a standard
deviation of 150~ms.

The distribution of initial magnetic magnetic fields is not well
known. \citet{vml+04} find that pulsars with $B_s > 2.5\times
10^{12}$~G account for more than half the pulsar birthrate and an
independent analysis by \citet{lfl+06} confirms these results. On the
other hand, Fig.~\ref{fg:ppdot} shows many pulsars exist with periods
of a few hundred milliseconds and $B_s$ in the range $10^{11}$ to
$10^{12}$~G. These pulsars have very probably been born close to their
present location in the $P-\dot P$ plane. Although their observed
number is comparable to the number of high-$B_s$ pulsars, their
birthrate is much lower because they move across the $P-\dot P$ plane
much more slowly.

\section{Optical Searches}
Early searches concentrated on optical wavelengths because X-ray
telescopes had insufficient senstivity and the expectation that the
surrounding nebula would be opaque at radio wavelengths. The first
Australian attempt was made by Bruce Peterson and the author on 26
February, 1987, just three days after the SN, using the 4-m
Anglo-Australian Telescope (AAT) with the dust cover nearly closed and
a high-speed photometer system previously used for pulsar
timing \citep{mp89}. In the next few months, several observations were
made using an 8-inch Celestron telescope on the ground beside the AAT
with the photometer system and then through 1988 at approximately
monthly intervals, again using the AAT. These searches gave a limit on
the pulsed fraction of the SN light of about $10^{-5}$ corresponding
to an I-magnitude limit of about 20 \citep{man88}. A similar search
was carried out at Las Campanas, Cerro Tololo and Mt Stromlo
observatories by \citet{pkm+89} with a similar upper limit on any
optical pulsations.

Then in March, 1989, \citet{kpm+89} announced the detection of a
periodicity at 0.508~ms based on a 7-hour observation with the 4-m
Cerro Tololo telescope. The periodicity had a sinusoidal variation
with a period of about 8 hours, suggesting the presence of a
Jupiter-sized planet around the pulsar. The very short pulse period
had major implications for neutron-star models and more than 50 papers
were published in the next year or so on the implications of this
result. Regretably, in 1990, it was realised that the periodicity
was spurious, due to interference from the TV guider on the
telescope \cite{kri91}. 

A Brazilian group claimed detection of optical pulsations
at 18.4~ms (reported by \citet{mur90}) but this was not
confirmed \citep{jab90}. An extensive search using various ESO
telescopes between 1988 March and 1990 April was reported
by \citet{oga+90} with limits of about 22nd magnitude for pulsations
at frequencies up to 5000 Hz.

\begin{figure}
\includegraphics[width=100mm]{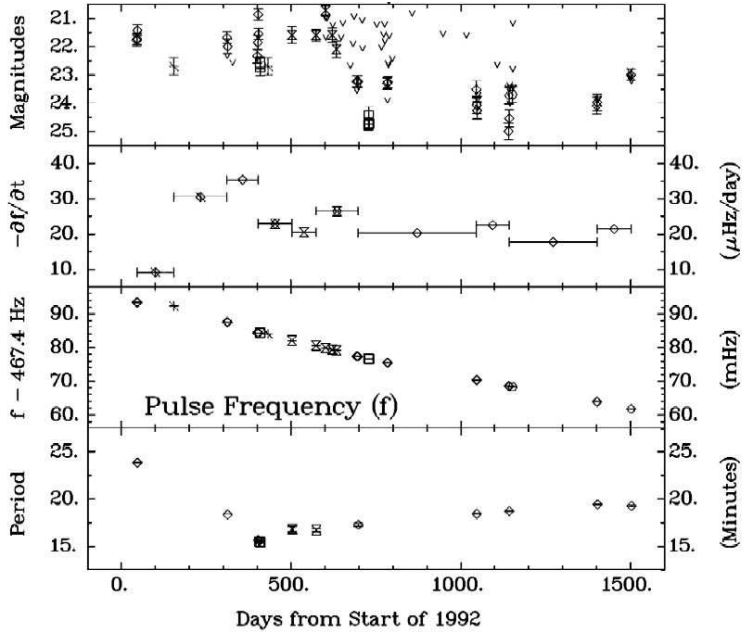}
\caption{Amplitude (in I magnitude) and frequency dependence of the 2.14~ms
pulsations detected by \citet{mkk+00} over four
years from 1992. The third panel from the top shows the variation of
pulse frequency, the second panel shows variations of the pulse
frequency derivative and the lowest panel shows variations of of the
modulation period. }\label{fg:mkk_freq} 
\end{figure}

In 1992 John Middledditch began circulating a series of preprints of
an extensive paper, ultimately published in 2000 \citep{mkk+00},
reporting the detection of optical pulsations at a period of 2.14~ms
using the Las Campanas 2.5-m Dupont Telescope and other
telescopes. Although there were numerous detections of a signal near
this period, the significance of each was relatively low and the
period and amplitude showed significant and largely unpredictable
fluctuations. In particular, a modulation of the signal with a period
of $\sim 1000$~s was observed. This modulation was of uncertain
origin, sometimes having the appearance of amplitude modulation and
sometimes frequency or phase modulation. Fig.~\ref{fg:mkk_freq} shows
the observed variations in amplitude of the signal, its fundamental
frequency and the frequency of the modulation. The amplitude of the
pulsed signal was highly variable with I magnitudes in the range 21 to
25 and a persistent spin-down with $d\nu/dt \sim 2\times
10^{-10}$~s$^{-2}$ was observed.
\citet{mkk+00} suggested that the observations were consistent with
precession and spin-down due to gravitational radiation from a neutron
star with a non-axisymmetric oblateness of $\sim 10^{-6}$. Despite
repeated observations by \citet{mkk+00} and others, no evidence for
this pulsation was found after 1996 and its identification with a
neutron star in SN 1987A remains doubtful.

An optical search using the High Speed Photometer system on the {\em
Hubble Space Telescope} was reported by \citet{pbb+95}. Four
observations in 1992 -- 1993 covering the band 160 -- 700 nm, each of
about 40~min duration and with 100 $\mu$s sampling, were searched with
both Fourier and time-domain folding methods. No significant
pulsations were observed in the period range 0.2~ms to 10~s with an
upper limit for the pulsed emission equivalent to a V magnitude of
$\sim 24$. A search with similar parameters was made by \citet{mp96}
using the 3.9-m Anglo-Australian Telescope. Data were obtained on two
nights of exceptional seeing and searched for periodicities in the
range 0.2~ms to 10~s. No signficant pulsations were observed with an
upper limit in V of about 24.6 magnitudes. 

\section{Radio Searches}
Most pulsars are detected at radio wavelengths and so, despite the low
probability of a detection at least in the first few years, searching
for a pulsar in SN 1987A at radio wavelengths was an obvious thing to
do. Early searches using the Parkes 64-m radio telescope at
frequencies between 400 and 5000 MHz were reported by \citet{man88}
with the best upper limit $\sim 0.2$~mJy at 1.5 GHz. Observations have
been made every 1 -- 2 years since then with similar parameters and
upper limits. 

A more extensive search was carried out at Parkes in 2006 December 19
-- 23 using three different receivers and different filterbank
systems. The frequencies and bandwidths used are listed in
Table~\ref{tb:radio_limits}. For all observations the sum of the power
in the two orthogonal polarisations for each frequency channel was
one-bit sampled and recorded on Digital Linear Tapes;
see \citet{mlc+01} for details of the data acquisition and recording
system.

\begin{table}
{\footnotesize
\caption{Results of a radio search for a pulsar in SN 1987A}\label{tb:radio_limits}
\begin{tabular}{ccccccccc}
\hline 
Band & Freq. & Total/Chan. & $t_{\rm int}$ & DM$_{\rm diag}$ & DM$_{\rm step}$ & DM Range & Time & Limit \\
& (MHz)  & BW (MHz)  & (ms) & (cm$^{-3}$ pc) & (cm$^{-3}$ pc) & (cm$^{-3}$ pc) & (h)  & ($\mu$Jy) \\ \hline
20cm & 1390  &  256/0.5  & 0.25 & 160  & 0.60 & 10--2420 & 2.3 & 115 \\
20cm & 1518  &  576/3.0  & 0.50 &70   & 0.68 & 10--2170 & 4.7 & 54  \\
10cm & 3083  &  864/3.0  & 0.25 &295  & 2.0  & 10--2100 & 4.7 & 90 \\
3cm &  8370  &  864/3.0  & 0.25 &5850 & 40   & 20--6020 & 4.7 & 58\\
\hline
\end{tabular}
}
\end{table}

The diagonal DM (DM$_{\rm diag}$) is the dispersion measure for which the dispersion
delay across one channel is equal to the sampling interval. For higher
dispersions the pulse is smeared and the sensitivity is reduced,
especially for short-period pulsars; for typical sensitivity curves as
a function of period and DM, see \citet{mlc+01}. The DM step (DM$_{\rm step}$) is 
the DM increment for a delay of two sample intervals across the full
bandwidth. For DMs up to DM$_{\rm diag}$ successive trial DM values
are separated by DM$_{\rm step}$. From DM$_{\rm diag}$ to 3~DM$_{\rm
diag}$ the trial DMs are separated by 2~DM$_{\rm step}$ and so on
until the maximum DM is reached. 

Data were processed using the SIGPROC pulsar search
package.\footnote{See \url{http://sigproc.sourceforge.net}.} Data were
first dedispersed at each trial DM and then compensated for the
Doppler shifts resulting from motion of the observatory relative to
the solar-system barycenter. The resulting time series was then
Fourier-transformed, harmonics summed and the power spectra
searched for candidates above a threshold of $7\sigma$. DM ranges
searched and observation times on SN 1987A are given in
Table~\ref{tb:radio_limits}. Transform lengths were $2^{24}$ samples
and the pulse period range searched was between the inverse Nyquist
frequency and 5~s. Observations of 2.3-h duration were also made with
each system at a position 30 arcmin south of the SN position to assist
with identifying spurious candidates.

No significant candidates with a S/N ratio greater than 9.0 were
observed in any system and in no case was there a convincing detection
of a genuine candidate in more than one receiver system or in
independent observations with the same system. The flux density limits
corresponding to a $9\sigma$ threshold in each analysis are given in
Table~\ref{tb:radio_limits}. In deriving these limits, the pulse duty
cycle was assumed to be 0.1.

\section{Other Limits on a Central Pulsar}
Many young pulsars drive a pulsar-wind nebula (PWN). The most famous
example is of course the Crab Nebula, but it is far from being
typical. Even of the known pulsars associated with SNR, only about
half have detectable PWN \citep{krh05}. With just a few exceptions,
all of the pulsars with associated PWN have spin-down luminosities
greater than $10^{36}$~ erg~s$^{-1}$. Only a minority of SNR have
detectable PWN. For example, the Cambridge SNR catalogue \citep{gre06}
contains 265 SNR of which only 39 have evidence for a PWN. Clearly,
detectable PWN are only associated with the most energetic young
pulsars.

Compact central objects (CCO) have been found in eight or nine
SNR \citep{pst04,pmk+06}. These objects have thermal X-ray spectra
consistent with surface emission from a neutron star. \citet{kfg+04}
searched for CCOs in a volume-limited sample of SNRs, placing
luminosity limits of $\sim 10^{31}$~erg~s$^{-1}$ on X-ray point
sources in four SNRs. Limits an order of magnitude higher are placed
on a further six SNR from the sample by \citet{kgks06}. These limits
are comparable to the lowest detected luminosities from CCOs, for
example, RX J0007.0+7302 in W44, which has a luminosity of $\sim
5\times 10^{31}$~erg~s$^{-1}$ \citep{szh+04}. Hence they do not
necessarily rule out the presence of a central neutron star in these
SNR.

Although SN 1987A, or more accurately, the resulting SNR, is getting
brighter at all wavelengths, imaging observations show that the
increasing emission is confined to the outer parts of the expanding
envelope and, at least in the optical and X-ray bands, primarily to
the region of interaction with the inner
ring \citep{gcc+05,bdd+06,pzb+04}. There is no evidence for a central
point (or near-point) source at any wavelength. \citet{slgs05} obtain
a luminosity limit in the 2 -- 10 keV band of $5\times
10^{34}$~erg~s$^{-1}$ using {\em XMM-Newton} data and \citet{pzb+04}
obtain a limit of $1.5\times 10^{34}$~erg~s$^{-1}$ using {\em
Chandra}. In the optical band from 290 to 965 nm \citet{gcc+05} give a
limit on the luminosity of any point source of $5\times
10^{33}$~erg~s$^{-1}$ using the {\em Hubble Space Telescope}.

These limits are well above luminosities of detected CCO in our Galaxy
and hence do not preclude the presence of a similar object in SN
1987A. Based on a bolometric luminosity limit of $3\times
10^{34}$~erg~s$^{-1}$ for a central PWN, \citet{oa04} find
that for a pulsar period of 300~ms, $B_s < 2.5\times
10^{12}$~G, not a very restrictive limit. 

At radio wavelengths, the best limit is obtained from the
9-GHz imaging where the expanding shell is best
resolved \citep{gsm+07}. However, even at this frequency, the
beamwidth is a signficant fraction of the nebular diameter and it is
difficult to limit the flux density of any central PWN or pulsar. A
conservative limit on the flux density of a central source is 1
mJy. If we assume a flat radio spectrum and a radio bandwidth of 20
GHz, then this corresponds to a radio luminosity of $\sim 3\times
10^{31}$~erg~s$^{-1}$. Let us assume that the entire spin-down energy
of a central pulsar is converted to luminosity of a surrounding
PWN. This is the most conservative assumption in the sense of giving a
lower limit to spin-down luminosity of any central pulsar. If we assume a spin
period of 200~ms, then for a spin-down luminosity of $3\times
10^{31}$~erg~s$^{-1}$, the required surface dipole magnetic field is
$B_s \sim 6\times 10^{10}$~G. As discussed above, these
parameters are within the range of possible birth parameters for
pulsars. If we make the more realistic assumption that only a fraction
of the spin-down luminosity goes to powering a PWN, then the pulsar
period could be shorter and/or the magnetic field stronger. 

It is abundantly clear that a central neutron star, if it exists, does
not have the high spin-down luminosity typical of pulsars associated
with Galactic PWN. However, it is also true that existing limits on
the luminosity of a central point source do not rule out the presence
of a perfectly plausible young neutron star with $P\gapp 100$~ms and
$B_s$ in the range $10^{11}$ to $10^{12}$~G at the centre of SN 1987A.

\section{Implications of the Pulsar Non-Detection}
Since the optical non-thermal emission from pulsars is evidently a
strong function of pulsar period \cite{ps87b}, optical searches are
doomed to failure if the central pulsar has a period of 100~ms or more
as suggested above.

\begin{figure}
\includegraphics[width=80mm,angle=270]{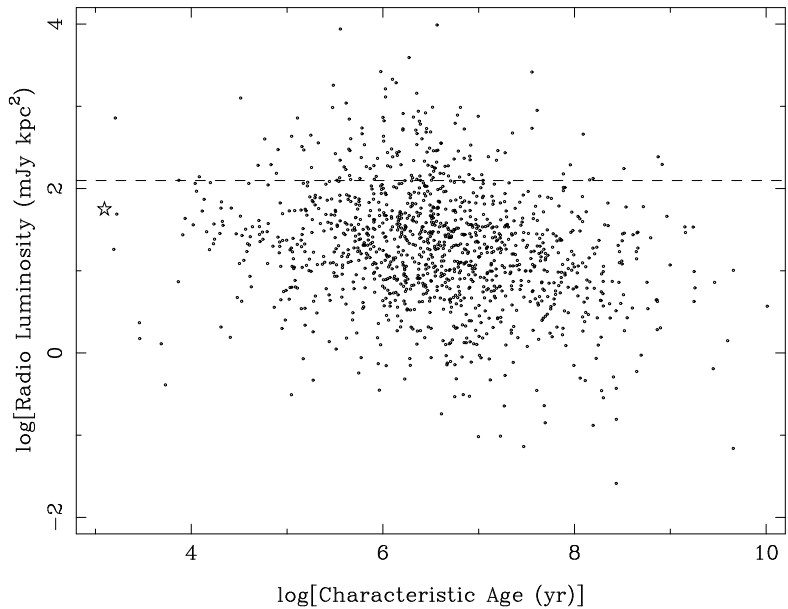}
\caption{Radio luminosity (expressed as $Sd^2$ where $S$ is the mean
flux density at 1400 MHz and $d$ is the pulsar distance) as a function
of characteristic age. The upper limit for a pulsar in SN 1987A is
shown by the horizontal dashed line and the Crab pulsar is marked by a
star. }\label{fg:rlum}
\end{figure}

The flux density limits given in Table~\ref{tb:radio_limits}
correspond to a radio luminosity (assuming an effective radiated
bandwidth of 1 GHz and a circular beam of width $36^\circ$) of about
$10^{21}$~erg~s$^{-1}$, much less than the likely spin-down luminosity
for a central pulsar. For a more direct comparison with the radio
luminosity of known pulsars, we use the psuedo-luminosity $L_r =
Sd^2$, where $S$ is the mean flux density at 1400 MHz and $d$ is the
pulsar distance. For $S=50\;\mu$Jy, the upper limit on $L_r$ is about
125~mJy~kpc$^2$. Fig.~\ref{fg:rlum} shows $L_r$ versus age for known
radio pulsars with this limit marked by a horizontal dashed
line. Because of the (relatively) large distance to the Magellanic
Clouds, the limit is not very restrictive. Even the Crab pulsar is
below the line and young but low-luminosity pulsars found in deep
searches toward SNR such as J1124-5916 in G292.0+1.8 \citep{cmg+02}
and J1930+1852 in G54.1+0.3 \citep{clb+02}, both of which have
$\tau_c \sim 2900$ yrs and $L_r \sim 2$~mJy kpc$^2$, are nearly two
orders of magnitude below it. Therefore, these radio searches
certainly do not exclude the presence of a low-luminosity pulsar
similar to many young pulsars in our Galaxy at the centre of SN 1987A.

Even if a pulsar with a high radio luminosity were present in SN
1987A, there are a number of reasons why we might not be able to
detect it. Most obviously, the pulsar may not be beamed toward us. The
beaming fraction (fraction of the celestial sphere swept over by the
beam as the star rotates) is not very well determined. Beaming
fractions are probably larger in the high-energy bands where the
emission is incoherent and could be 50\% or more. In the radio
band, a beaming fraction of 20\% is often assumed, but for young
pulsars it could be considerably larger. Radio beams are generally
quite patchy, so even if the over-all beamwidth is large, there is a
finite chance of being missed by the stronger parts of the beam. 

It is possible that the pulsar magnetic field takes some time to
develop. Most current models assume either that the field a compressed
stellar field or that it is generated during the collapse by dynamo
action \citep[e.g.,][]{bub05}, but other models
exist \citep{br88} in which the field growth takes place
after the neutron-star formation. If the growth timescale is decades, then
even a rapidly spinning neutron star could still be undetectable. 

Even though the outer parts of the SN 1987A nebula are now quite
transparent in all relevant bands, it is possible that the immediate
environment of the central star still has a relatively high gas
density which would cause scattering and/or absorption of emission
from the star. \citet{fcp99} discuss fallback of
stellar ejecta on to the neutron star and show that it will have a
very high opacity at early times and will be largely driven off by radiation
pressure. However, fallback is possible decades after the SN
explosion \cite{ww95} and this may still form a relatively dense and
turbulent nebula surrounding the star. 

Finally, there is the possibility that a neutron star formed at the
time of the SN accreted so much matter from fallback of ejecta that it
exceeded the maximum mass of a neutron star and became a black
hole \cite{bb95,che96}. \citet{fcp99} argue that this is unlikely, or at
least subject to a restrictive range of conditions. 

\section{Conclusions}
The enormously exciting prospect of being able to study a pulsar in
its earliest years has motivated a large number of searches for a
pulsar in SN 1987A. These searches were primarily at optical and radio
wavelengths since our experience with other young pulsars shows that
these are the most likely bands for a successful
detection. Regretably, despite all this effort (and several false
alarms) no pulsar has been detected. Furthermore, limits on the
luminosity of a central point (or near point) source at the centre of
SN 1987A show that any central pulsar is not highly energetic like the
Crab pulsar or PSR B0540$-$69.

However, all of these limits leave open a region of parameter space
for a young neutron star in the centre of SN 1987A. Specifically, a
neutron star with a rotation period $P\gapp 100$~ms and a surface
dipole magnetic field strength $B_s$ in the range $10^{11}$ to
$10^{12}$~G is not ruled out by any observations so far. There is
clear evidence that some neutron stars are born in the Galaxy with
parameters in this range.

Therefore, continued searches for a pulsar in SN 1987A are certainly
justified. The greatly increased sensitivity of the proposed Square
Kilometer Array will be highly beneficial to radio searches since the
current luminosity limits are well above typical luminosities of young
radio pulsars. If the pulse period is relatively long, detection of
non-thermal optical or X-ray pulsations is unlikely, but more senstive
searches in the future may detect thermal emission from the
neutron-star surface and possibly modulation due to rotation of the
star.


\begin{theacknowledgments}
I thank Chris Lustri for his assistance with analysis of the data from
the 2006 December Parkes observations, John Reynolds for his
invaluable assistance with configuring the Parkes receiver systems and
Bryan Gaensler for data on estimated pulsar birth periods. The Parkes radio
telescope is part of the Australia Telescope which is funded by the
Commonwealth Government for operation as a National Facility managed
by CSIRO.
\end{theacknowledgments}




\end{document}